\documentstyle[aps,preprint,eqsecnum]{revtex}
\tightenlines
\begin{document}

\title{Classical mappings of the symplectic model
and their application to the theory of large-amplitude
collective motion}
\author{Abraham Klein and Niels R.\ Walet\thanks{address after Sept.~1, 1993:
Instit\"ut f\"ur theoretische Physik III, Universit\"at Erlangen-N\"urnberg,
D-91058 Erlangen, Germany}}
\address{Department of Physics, University of Pennsylvania,
Philadelphia, Pennsylvania 19104-6396, USA}

\maketitle

\begin{abstract}
We study the algebra Sp(n,R) of the symplectic model,
in particular for the cases n=1,2,3,
in a new way.  Starting from the Poisson-bracket realization
we derive a set of partial differential equations for the
generators as functions of classical canonical variables. We obtain
a solution to these equations that represents the
classical limit of a boson mapping of the algebra.
We show further that this mapping plays
a fundamental role in the collective description of many-fermion systems
whose Hamiltonian may be approximated by polynomials in the associated
algebra, as is done in the simplest versions of the symplectic model.
The relationship to the  collective dynamics is formulated as a theorem
that associates the mapping with an exact solution of the time-dependent
Hartree approximation.   This solution determines a decoupled classical
symplectic manifold, thus satisfying the criteria that define an exactly
solvable model in the theory
of large amplitude collective motion.  The models thus obtained also
provide a test of methods
for constructing an approximately decoupled manifold
in fully realistic cases.  We show that an algorithm developed in
one of our earlier works reproduces the main results of the theorem.
\end{abstract}


\narrowtext

\section{Introduction}

We have been engaged for a decade in an effort to formulate
a theory of large amplitude collective motion with the special aim
of applying it to nuclear physics.  The theory has both a classical and
a quantum dimension.  The classical aspect has been most fully developed
and described in a review \cite{CCM}.  The quantum aspect is presently in a
stage of vigorous development \cite{QM1,QM2}, supplanting the early work
on this part of the theory \cite{QCM1,QCM2}.  At the same time a program
of applications to
problems of nuclear structure has been undertaken \cite{ACM1,ACM2,ACM3,ACM4}.

Early in the
latter work, we became aware of a paucity of solvable models with some
physical content.  The usefulness of such models is that they provide a
testing ground for the algorithms that would later be applied to more
realistic models. For our initial investigation
we selected a well-known model of monopole
vibrations \cite{MM1}, exactly solvable because the Hamiltonian is a polynomial
in the generators of the algebra Sp(1,R) (or SU(1,1)). We studied this
model in two ways.  First, by utilizing the classical limit of the algebra,
we were able to produce an exact solution of the time-dependent Hartree
equations (derived previously by rather less transparent techniques
\cite{MM2})
and by means of this solution a decoupled collective Hamiltonian for the
monopole vibration.  Second, and more important, we could check if the
same Hamiltonian emerged from the application of the theory of large
amplitude collective motion. The monopole model
provided us with
an apparently ideal test of the soundness of our algorithms.
This test failed initially, forcing us eventually to recognize and correct an
incompleteness
in our previous theory.

The first goal of this
paper is to show that the method developed for the algebra of Sp(1,R)
can be extended to the algebra of Sp(n,R).  In particular, we work out fully
the cases n=2 and n=3.  The former leads to a Hamiltonian with three collective
coordinates, describing the interaction of a monopole degree of freedom
with a quadrupole tensor in two spatial coordinates and is thus only of
interest as a toy model.  On the other hand n=3 leads to a Hamiltonian with
six degrees of freedom describing both a monopole and a three-dimensional
quadrupole.  This defines the model as not only one of physical
interest {\em per se}, but also because of its connection with the
symplectic \cite{SM} and pseudo-symplectic \cite{PSM} models. The latter, in
particular, provides a possibly useful truncation scheme for shell-model
calculations for other than the lightest deformed nuclei.

Though not one
of the aims of the present paper, this identification will allow us
on a future occasion to compare
the quantum consequences of the collective Hamiltonian to be derived in this
paper with the results of an exact diagonalization carried out for the
original many body Hamiltonian.
In fact such a comparison has broader implications than the accuracy achieved
for the special Hamiltonian considered, since it has been demonstated
that Hamiltonians consisting of suitably chosen polynomials in the generators
can give a rather precise fit to the low-energy spectra and other properties
of even deformed nuclei \cite{PSM}.

The second aim of this paper is to demonstrate that the extended algorithm
formulated in connection with the monopole model also provides correct
results for the generalized models.

The presentation is organized as follows:  In Sec.\ 2 we give a brief
summary of the properties of the algebra of Sp(n,R) needed in the ensueing
development as well as a discussion of the model Hamiltonians to be studied.
In the following three sections we then study separately the cases n=1,2,3.
We describe in each instance the mapping  of the
algebra onto a classical symplectic manifold, which is tied to
the existence of a manifold of  solutions of the time-dependent Hartree
equation and an associated decoupled collective Hamiltonian.
We then show how the
same collective Hamiltonian can be derived from our theory of large-amplitude
collective motion.  The material for the monopole case is a rearrangement
with different emphasis of results presented previously \cite{ACM1}.  All other
results are new.  In Sec.\ 6, we make suggestions
for further work, involving both applications and extensions of the results
of this paper, as well as the study of possible connections with previous
research.  Two appendices, A and C contain important details of
the calculations that would impede the flow of the argument in the main
text.  In Appendix B we review briefly the generalization of the theory
suggested by our previous study of the monopole model and indicate how it
applies to the more general cases.

\section{Algebraic and dynamical preliminaries \label{sec:ADP}}
\subsection{The algebra of Sp(n,R)}

The defining algebra of the group Sp(n,R) is given most simply in terms
of $n$ boson pairs, $a_i$, $a^i$, $i=1\cdots n$, where the subscript
identifies the destruction and the superscript the creation operators.
One standard set of generators is
composed of three distinct bilinear forms in these
operators,       \begin{eqnarray}
A_{ij} &=& a_i a_j,  \label{eq:2.1}   \\
A^{ij} &=& a^i a^j,  \label{eq:2.2}   \\
B^i_j &=& a^i a_j +\mbox{$\frac{1}{2}$}\delta^i_j.  \label{eq:2.3}
\end{eqnarray}

An equivalent set of generators, more useful for the purposes of this paper
is given in terms of single-particle coordinate and momentum operators,
$x_i$ and $p_i$, respectively,
related to the boson operators in the standard way, namely,
\begin{eqnarray}
a_i &=& \frac{1}{\sqrt{2}}(x_i + ip_i),  \label{eq:2.9} \\
a^i &=& \frac{1}{\sqrt{2}}(x_i - ip_i).  \label{eq:2.10}   \end{eqnarray}
The alternative set of generators takes the form         \begin{eqnarray}
Q_{ij} &=&  x_i x_j, \label{eq:2.11}  \\
K_{ij} &=& p_i p_j, \label{eq:2.12}  \\
L_{ij} &=& \mbox{$\frac{1}{2}$} (x_i p_j - x_j p_i), \label{eq:2.13} \\
S_{ij} &=& \mbox{$\frac{1}{2}$} (x_i p_j + x_j p_i) -\mbox{$\frac{1}{2}$}
i\delta_{ij},  \label{eq:2.14}
\end{eqnarray}
satisfying the set of commutation relations,
\mediumtext
\begin{eqnarray}
[Q_{ij},K_{kl}]&=&i(\delta_{jk}S_{il}+\delta_{jl}S_{ik}+\delta_{ik}S_{jl}
+\delta_{il}S_{jk})  \nonumber \\
&& +i
(\delta_{jk}L_{il}+\delta_{jl}L_{ik}+\delta_{ik}L_{jl}+\delta_{il}L_{jk}),
\label{eq:2.15} \\
{}[Q_{ij},L_{kl}]&=&\mbox{$\frac{1}{2}$}
i(\delta_{lj}Q_{ik}-\delta_{jk}Q_{il}+\delta_{il}Q_{jk}
-\delta_{ik}Q_{jl}), \label{eq:2.16}   \\
{}[Q_{ij},S_{kl}]&=&\mbox{$\frac{1}{2}$}
i(\delta_{lj}Q_{ik}+\delta_{jk}Q_{il}+\delta_{il}Q_{jk}
+\delta_{ik}Q_{jl}), \label{eq:2.17}   \\
{}[K_{ij},L_{kl}]&=&\mbox{$\frac{1}{2}$}
i(\delta_{il}K_{jk}+\delta_{jl}K_{ik}-\delta_{ik}K_{jl}
-\delta_{jk}K_{il}),  \label{eq:2.18} \\
{}[K_{ij},S_{kl}]&=&-\mbox{$\frac{1}{2}$}
i(\delta_{il}K_{jk}+\delta_{jl}K_{ik}+\delta_{ik}K_{jl}
+\delta_{jk}K_{il}),  \label{eq:2.19} \\
{}[L_{ij},S_{kl}]&=&\mbox{$\frac{1}{2}$}
i(\delta_{ik}S_{jl}-\delta_{jl}S_{ik}+\delta_{il}S_{jk}
-\delta_{jk}S_{il}), \label{eq:2.20}  \\
{}[L_{ij},L_{kl}]&=&\mbox{$\frac{1}{2}$}
i(\delta_{ik}L_{jl}+\delta_{jl}L_{ik}-\delta_{il}L_{jk}
-\delta_{jk}L_{il}), \label{eq:2.21}  \\
{}[S_{ij},S_{kl}]&=&-\mbox{$\frac{1}{2}$}
i(\delta_{ik}L_{jl}+\delta_{jl}L_{ik}+\delta_{il}L_{jk}
+\delta_{jk}L_{il}). \label{eq:2.22}  \end{eqnarray}
\narrowtext
In terms of these generators, the second order Casimir invariant is given by
the expression   \begin{equation}
{\cal C}_2 = -\mbox{$\frac{1}{2}$} \{Q_{ij},K_{ij}\} +S_{ij}S_{ij}
-L_{ij}L_{ij}.
\label{eq:2.23}  \end{equation}

\subsection{Dynamical preliminaries}

In the body of the paper we shall find it convenient to assign a different
meaning to the upper-case symbols used above to denote the various generators.
We shall therefore refer to these quantities
in the text below by the corresponding
lower-case letters, $Q_{ij}\rightarrow q_{ij}$, etc.   We shall be
concerned with a many-body realization of the algebra, for which we shall
use the second-quantized formalism.  Thus, if $\psi({\bf x})$, $\psi^{\dag}
({\bf x}')$ satisfy the anticommutation relations,   \begin{equation}
\{\psi({\bf x}),\psi^{\dag}({\bf x}')\} =
\delta({\bf x} -{\bf x}'), \label{eq:2.24}
\end{equation}  where ${\bf x}$ stands for the $n$-dimensional vector
$(x_1\cdots x_n)$ and any additional intrinsic variables such as spin
and isospin that are included in the dynamics,  then the operators
$\hat{Q}_{ij}$, $\hat{K}_{ij}$, etc., where, for example,  \begin{equation}
\hat{Q}_{ij} = \int\,d{\bf x}\psi^{\dag}({\bf x})q_{ij}\psi({\bf x}),
\label{eq:2.25a}  \end{equation}
are the set of generators of Sp(n,R) relevant to the fermion many-body
problem.  Below we shall also suppress the boldface for the quantity
${\bf x}$.

We shall be concerned further with studying the classical limit
of Hamiltonian operators belonging to the enveloping algebra of Sp(n,R)
for n=1,2,3, i.e., Hamiltonians that are polynomials in the
generators of these algebras.
Consider the one-dimensional
case, where we have just three generators, $\hat{Q}_{11}$, $\hat{K}_{11}$,
and $\hat{S}_{11}$, that we rename by dropping the subscripts.  We then
study a Hamiltonian, \begin{equation}
\hat{H} = \mbox{$\frac{1}{2}$} (\hat{K}+\hat{Q}) + \mbox{$\frac{1}{2}$} \kappa
\hat{Q}\hat{Q},
\label{eq:2.25}   \end{equation}
that is the sum of an harmonic oscillator part and of a ``monopole-monopole"
interaction. The considerations to be applied to this Hamiltonian can
be generalized to one in which we add any polynomial in the operator
$\hat{Q}$, but we shall quote detailed results only for the  Hamiltonian
(\ref{eq:2.25}).

Consider an arbitrary Slater determinant
describing $N$ fermions. We shall
be interested in the expectation value of (\ref{eq:2.25}) in this state,
evaluated in Hartree approximation, i.\ e.\, to the leading order in $N$,
the number of particles.
Using angular brackets to denote this average, we have in this approximation
\begin{equation}
\langle \hat{H}\rangle \equiv H_C = \mbox{$\frac{1}{2}$} (K+Q)
+\mbox{$\frac{1}{2}$}\kappa Q^2,
\label{eq:2.26}    \end{equation}
where, e.\ g.\ ,        \begin{equation}
Q = \langle \hat{Q}\rangle,    \label{eq:2.27}   \end{equation}
i.\ e.\ ,  the Hartree average of any generator will be denoted by the
same symbol without hat.  As a further instance, the Hartree approximation
to the Casimir invariant of Sp(1,R) is    \begin{equation}
\langle{\cal C}_2\rangle ={\cal C}_2 = -QK + S^2. \label{eq:2.28}
\end{equation}

The most important idea that informs the next section is
that we consider $Q$ to be a classical
collective coordinate and associate it with a corresponding canonical
momentum, $P$. The Hartree approximation to the generators
then defines a classical
limit of the algebra in which the commutators are replaced by Poisson brackets.
These are a set of partial differential equations for the classical generators,
$Q$,$K$, and $S$, that determine the latter
as functions of $Q$ and $P$.  This mapping will not only determine the
Hartree average as a function of $Q$ and $P$, but will allow us to associate
this classical boson mapping with a solution of the time-dependent
Hartree equation.  The concepts mentioned here will be rendered precise in
the next section and then extended to two and three dimensions in the following
sections.

\section{The monopole model associated with the
algebra of Sp(1,R) \label{sec:monopole}}
\subsection{Classical mapping and Hartree solution}

As stated in the introduction, the results to be presented in this section
have appeared in our previous work \cite{ACM1}.  What we aim for here is a more
systematic presentation with enhanced emphasis on the significance of the
results.  We do this by collecting the results into a theorem.    \\
{\em Theorem}:
For a class of many-Fermion Hamiltonians belonging to the enveloping algebra
of Sp(1,R),
of which Eq.\ (\ref{eq:2.25})
is a prototype, there exists a two parameter family of Slater determinants,
defined by density matrices, $\rho(xx'|Q,P)$, that describe states
belonging to an irreducible representation. These states   \\
(i) induce a mapping of Sp(1,R) onto a symplectic  manifold $(Q,P)$ in which
the generators, $\hat{G}$, are mapped as classical dynamical
variables, i.e.,   \begin{equation}
\hat{G} \rightarrow G(Q,P),      \label{eq:2.29}   \end{equation}
and in particular,   \begin{equation}
\hat{Q} \rightarrow Q. \label{eq:2.30}  \end{equation}
The Hartree expectation value of any member of the enveloping
algebra is thereby also
mapped.  \\
(ii) The associated density matrices have the diagonal form   \begin{equation}
\rho(x,x'|Q,P) =\sum_h \psi_h (x|Q,P)\psi_h^{\ast}(x'|Q,P),  \label{eq:2.31}
\end{equation}    where the sum is over the $N$ occupied orbitals, $h$.
Furthermore $\psi_h (x|Q,P)$ can be written as a product     \begin {equation}
\psi_h (x|Q,P) = \exp(iPx^2)\phi_h (x,Q),  \label{eq:2.32}  \end{equation}
and $\phi_h$ is the solution of a constrained Hartree equation
\begin{eqnarray}
\epsilon_h \phi_h &=& ({\cal H}-\lambda x^2)\phi_h, \label{eq:2.33}  \\
{\cal H} &=& \mbox{$\frac{1}{2}$} (p^2 + x^2) +\kappa Qx^2,   \label{eq:2.34}
\\
\lambda &=& \frac{dH_C (Q,P=0)}{dQ} \equiv \frac{dV}{dQ}.  \label{eq:2.35}
\end{eqnarray}
Here ${\cal H}$ is the Hartree Hamiltonian and $H_C$ is the classical
collective Hamiltonian defined in (\ref{eq:2.26}).
These results identify the density matrix (\ref{eq:2.31}) as a solution
of the time-dependent Hartree equation   \begin{equation}
i\dot{\rho} = [{\cal H}, \rho].    \label{eq:2.36}   \end{equation}

We turn to the proof of part (i) of the theorem.  The Hartree average
of a product of two generators, $\hat{G}_1$ and $\hat{G}_2$,  \begin{equation}
\langle \hat{G}_1\hat{G}_2\rangle =  G_1(Q,P)G_2(Q,P)  \label{eq:2.37}
\end{equation}    can be identified as the leading term in
the convolution of these two operators under a Wigner transform with
respect to the collective variables $Q,P$.
For a consistent evaluation of commutators, however,
we need the next term in the convolution,  \begin{equation}
\langle [\hat{G}_1,\hat{G}_2]\rangle \rightarrow i[G_1 (Q,P),G_2 (Q,P)]_{PB},
\label{eq:2.38}  \end{equation}  where the Poisson bracket is to
be evaluated with respect to the single canonical pair (Q,P).
{}From Eqs.\ (\ref{eq:2.15})-(\ref{eq:2.22}),
we thereby obtain the Poisson bracket algebra
\begin{eqnarray}
[Q,S] &=& \frac{\partial S}{\partial Q} = 2Q,   \label{eq:2.39}  \\
{}[Q,K] &=& \frac{\partial K}{\partial Q} = 4S , \label{eq:2.40}  \\
{}[K,S] &=& -2K. \label{eq:2.41}    \end{eqnarray}

Treated in turn, these  differential equations yield
the relations \begin{eqnarray}
S &=& 2QP, \label{eq:2.42}   \\
K &=& 4QP^2 + \chi(Q),\;\;\;  {\rm with}  \label{eq:2.43} \\
Q\frac{d\chi}{dQ} &=& -\chi(Q).   \label{eq:2.44}    \end{eqnarray}
Thus   \begin{equation}
\chi(Q) = C/Q.    \label{eq:2.45} \end{equation}
The constant $C$ can be obtained by evaluating the second-order Casimir
invariant in the Hartree approximation.   We find \begin{equation}
\langle \hat{{\cal C}}_2\rangle = -C = -N^4/4,  \label{eq:2.46} \end{equation}
where the pair of equalities express the results of two separate procedures.
On the one hand the value $-C$ is obtained by direct substitution of the mapped
generators.  On the other hand the specific value $-N^4/4$ is obtained by
calculating the Casimir invariant for a simple state in the irreducible
representation, as explained in Appendix A.
Finally, we record that the collective Hamiltonian, which is the Hartree value
of the many-particle Hamiltonian, maps to  \begin{eqnarray}
H_C &=& 2QP^2 + \mbox{$\frac{1}{2}$} Q +\mbox{$\frac{1}{2}$} \kappa Q^2
+N^4/8Q,  \nonumber  \\
  & \equiv & 2QP^2 +V(Q),  \label{eq:2.47}     \end{eqnarray}
  and we have also displayed the collective potential energy.  Note, however,
that the singular term in the latter,
that has a tantalizing resemblance to the scalar Berry potential \cite{QM1},
originates in the many-particle kinetic energy.

We turn next to part (ii) of the theorem, which provides a construction of
the manifold of density matrices associated with the classical mapping
just given.  We show first that the single-particle wave functions of which
the density matrix is composed are of the form  \begin{equation}
\psi_h (x|Q,P) =\exp(i{\cal S}(x,P))\phi_h(x,Q),  \label{eq:2.48}
\end{equation}  i.\ e,\, the dependence on the collective momentum is
independent of the orbit label.  To see this, let us calculate the time
derivative of the classical variable $Q$ in two ways, directly from the
classical equations of motion,   \begin{equation}
\dot{Q} = \frac{\partial H_C}{\partial P} = 4PQ, \label{eq:2.49}
\end{equation} and by actually evaluating the Hartree approximation of the
quantum equations of motion,   \begin{eqnarray}
\dot{Q} &=& -i\langle [\hat{Q},\hat{H}]\rangle =2\langle\hat{S}\rangle
\nonumber \\
&=& 2\int dx\,x\frac{\partial{\cal S}}{\partial x}\rho(x,x).  \label{eq:2.50}
\end{eqnarray}  In obtaining this last result, we have assumed that the
orbitals $\phi_h$ are real.  Equations (\ref{eq:2.49}) and (\ref{eq:2.50})
yield the solution    \begin{equation}
{\cal S} =Px^2.     \label{eq:2.51}   \end{equation}

To find the orbitals $\phi_h$ introduced in Eq.\ (\ref{eq:2.48}),
we back up a notch by defining
\begin{equation}
\tilde{\psi}_h =\exp\{-i\int_0^t dt'\, \epsilon_h [Q(t')]\} \psi_h.
\label{eq:2.52}   \end{equation}
The extra adiabatic phase in (\ref{eq:2.52}) drops out of the density matrix,
but must be included in order that
$\tilde{\psi}$ satisfy the time-dependent Hartree equation,
\begin{equation}
i\frac{d\tilde{\psi}}{dt} = {\cal H}\tilde{\psi}.  \label{eq:2.53}
\end{equation}
Taking into account both the explicit and the implicit
time dependence contained in (\ref{eq:2.53}), the latter is replaced by the
equation   \begin{equation}
\epsilon_h\psi_h + i\dot{P}\frac{\partial \psi_h}{\partial P}
+i\dot{Q}\frac{\partial \psi_h}{\partial Q} ={\cal H}\psi_h.
\label{eq:2.54}    \end{equation}       Substituting the classical equations
of motion for the time derivatives and inserting the
form of $\psi_h$, the resulting equation has terms of zero, first,
and second order in the momentum $P$.  The terms of second order are found
to cancel, whereas the terms of zero and first order give, respectively, the
equations     \begin{eqnarray}
\epsilon_h \phi_h &=& ({\cal H} -\lambda x^2)\phi_h, \label{eq:2.55}  \\
4Q\frac{\partial \phi_h}{\partial Q} &=& -\phi -2x\frac{\partial \phi}
{\partial x}.  \label{eq:2.56}     \end{eqnarray}

We deal first with (\ref{eq:2.55}).  By means of the defintions
\begin{eqnarray}
\bar{{\cal H}} &=& \mbox{$\frac{1}{2}$} (p^2 +\bar{\omega}^2 x^2 ),
\label{eq:3.29} \\
\bar{\omega}^2 &=& 1 + 2(\kappa Q -\lambda),  \label{eq:3.30}   \end{eqnarray}
(\ref{eq:2.55}) becomes    \begin{equation}
\epsilon_h \phi_h = \bar{{\cal H}}\phi_h,  \label{eq:3.31}
\end{equation}   with the normalized solution, in terms of solutions
$\phi_h^{(sho)}$ for the simple harmonic oscillator with unit mass and
unit frequency,  \begin{equation}
\phi_h (x) =\bar{\omega}^{1/4}\phi_h^{(sho)}(\sqrt{\bar{\omega}}x).
\label{eq:3.32}   \end{equation}
It remains only to verify (\ref{eq:2.56}).  This can be done, using
the explicit form of $\bar{\omega}^2$, derived from Eq.\ (\ref{eq:3.30})
by substituting the value of $\lambda=(dV/dQ)$ from Eq.\ (\ref{eq:2.47}),
namely,   \begin{equation}
\bar{\omega}^2 =C/Q^2.     \label{eq:3.33}    \end{equation}
This completes the proof of the theorem.

\subsection{Application of the theory of large amplitude collective motion}

It is not the purpose of this section to review yet again the theory of
large amplitude collective motion.  It  has been described exhaustively
in the works referred to in the introduction.
In so far as the present discussion is concerned
the principal aim is to find the collective Hamiltonian from this theory.
The first step of the procedure is to assume that the collective variable
is the expectation value of a one-body operator.
For the model Hamiltonian, Eq.\ (\ref{eq:2.25}),
with a separable interaction, the
logical starting choice is always determined by the ingredients of
that interaction.  In
the present instance, the choice is the Hartree expectation value of
$\hat{Q}$, or in other words it is determined by the mapping
$\hat{Q}\rightarrow Q$. This leads automatically
to the constrained Hartree equation
solved in part (ii) of the theorem.

The next step is to compute the
many-particle Hartree energy associated with the filling of these orbits
with $N$ particles.  We outline the calculation, in which this Hartree
energy is identified as the potential energy, $V(Q)$, of the system.
We have first  \begin{equation}
V(Q)\equiv \langle\hat{H}\rangle = {\rm tr}[\mbox{$\frac{1}{2}$} (p^2 +x^2
)\rho ]
+\mbox{$\frac{1}{2}$}\kappa Q^2.  \label{eq:3.35}     \end{equation}
This equation can be transformed into a more useful version, as follows:
We first calculate the sum of the single-particle energies, \begin{eqnarray}
\sum_h\epsilon_h &=& {\rm tr}(\bar{{\cal H}}\rho) \nonumber \\
&=& {\rm tr}[\mbox{$\frac{1}{2}$} (p^2 + x^2 )\rho] +\kappa Q^2 -Q(dV/dQ).
\label{eq:3.35a}
\end{eqnarray}    We then invoke
the virial theorem, in the form  \begin{equation}
\sum_h\epsilon_h =\bar{\omega}^2{\rm tr}(x^2\rho) =\bar{\omega}^2 Q,
\label{eq:3.36}   \end{equation}
which allows us to replace (\ref{eq:3.35a}) by the expression
  \begin{equation}
{\rm tr}[\mbox{$\frac{1}{2}$} (p^2 +x^2 )\rho] =\bar{\omega}^2 Q -\kappa Q^2
+Q\frac{dV}{dQ}.
\label{eq:3.37}     \end{equation}
When the latter is now substituted into (\ref{eq:3.35}),
the result is a differential equation for the potential energy,
\begin{equation}V(Q) =Q +\frac{3}{2}\kappa Q^2 -Q\frac{dV}{dQ}, \label{eq:3.38}
\end{equation}
of which the solution is   \begin{equation}
V(Q) = \mbox{$\frac{1}{2}$} Q + \mbox{$\frac{1}{2}$}\kappa Q^2 +C/Q.
\label{eq:3.39}  \end{equation}
The last term is the solution of the homogeneous version of (\ref{eq:3.38}),
depending on the same constant that appeared in Eq.\ (\ref{eq:2.45}).

The next stage in this procedure is to calculate the collective mass, $B$,
defined by writing the collective kinetic energy in the form
$\mbox{$\frac{1}{2}$} BP^2$.
We shall calculate $B$ in two (equivalent) ways.  The first, that we have not
used in our previous work is connected to the work done in proving the
theorem of the previous section.  Its deceptive simplicity carries special
instruction to which we shall return below.  The procedure is to make use
once more of the two values of $\dot{Q}$ found in Eqs.\
(\ref{eq:2.49}) and (\ref{eq:2.50}),
without assuming that we know the coefficient of $P^2$.  This yields the
equality  \begin{equation}
BP = 2\langle\hat{S}\rangle =2S.    \label{eq:3.40}  \end{equation}
Since we have already solved for the value of $S$ from the PB form of the
algebra, we may substitute from (\ref{eq:2.42}) into (\ref{eq:3.40})
and verify the
value $B=4Q$ found previously.  (We are not depending on the theorem.  If
we didn't know $S$ as a classical dynamical variable, we would calculate
it now.)  .
According to the general theory, the fact
that $B$ depends only on $Q$ verifies that we are dealing with an exactly
decoupled mode.  This means that we have chosen the collective coordinate
correctly, i.e., self consistently.

We shall next attempt to convince the reader that we have not engaged in
sleight of hand.  We can do this by complicating the model in what appears
to be a minimal way, but taking it outside of the algebra Sp(1,R).  We
do this by adding to $\hat{H}$ a term $\mbox{$\frac{1}{2}$}\kappa_4
\hat{Q}_4^2$, where
$\hat{Q}_4$ is the many body version of the operater $q_4 =x^4$.  Even
if we crank only with $x^2$, we now find that the {\em potential}
energy is a function both of $Q$ and of $Q_4$, so that the minimum number
of collective variables that we can introduce is two.  We must therefore
apply the procedure followed above to calculate the mass to the
two-dimensional domain with kinetic energy, $T$, given by     \begin{equation}
T =\mbox{$\frac{1}{2}$} B^{00}P^2 +B^{04}PP_4 +\mbox{$\frac{1}{2}$} B^{44}P_4
P_4.   \label{eq:3.41}
\end{equation}   We shall not carry this through in detail, but, in
fact, it is a perfectly straightforward exercise and yields the mass matrix
\begin{eqnarray}
B^{00} &=& 4Q,  \label{eq:3.42}  \\
B^{04} &=& 8Q_4 ,  \label{eq:3.43}  \\
B^{44} &=& 16Q_6,  \label{eq:3.44}    \end{eqnarray}
where $Q_6$ is the Hartree average of $x^6$.  The point
is that in the classical collective Hamiltonian, $Q_6$
is independent of the two variables introduced previously.  Thus the
effort to decouple a finite symplectic manifold has failed.  There is no
cure for this ailment.  Thus we might try to introduce $Q_6$ as an additional
collective variable by cranking on $x^6$ and thus bringing it into the
potential energy.  By extension with what has been found above, however,
the augmented kinetic energy will bring in still higher powers of $x$.

The method of calculating the collective mass developed in our previous
work \cite{ACM1} that gives the same result as above
is described in Appendix B.

\section{Monopole plus two-dimensional model associated with the algebra
of Sp(2,R)}
\subsection{Classical mapping and Hartree solution}

We turn to the problem of stating and proving a theorem that generalizes
the results of the previous section.  We first specify the model to be studied.
Here it is convenient to replace the coordinate
generators, $\hat{Q}_{ij}$ by a monopole $\hat{Q}_0$, and a two-dimensional
quadrupole, $\hat{Q}_i,\;i=1,2$, where  \begin{eqnarray}
\hat{Q}_0 &=& \int \psi^{\dag}(x)\mbox{$\frac{1}{2}$} (x_1^2 +x_2^2)\psi(x),
\label{eq:4.1} \\
\hat{Q}_1 &=& \int \psi^{\dag}(x)\mbox{$\frac{1}{2}$} (x_1^2 -x_2^2)\psi(x),
\label{eq:4.2}  \\
\hat{Q}_2 &=& \int \psi^{\dag}(x)x_1 x_2\psi(x).
\label{eq:4.3}    \end{eqnarray}
The Hartree expectation values of these operators will be interpreted as
their classical maps.  We consider one further definition, namely,
\begin{equation}
\hat{K}_0 = \mbox{$\frac{1}{2}$} (\hat{K}_{11} + \hat{K}_{22}).  \label{eq:4.4}
\end{equation}
As model Hamiltonian we choose   \begin{equation}
\hat{H} = \mbox{$\frac{1}{2}$} (\hat{K}_0 + \hat{Q}_0) +\mbox{$\frac{1}{2}$}
\kappa_0 \hat{Q}_0\hat{Q}_0
+ \mbox{$\frac{1}{2}$} \kappa_2 (\hat{Q}_1\hat{Q}_1 +\hat{Q}_2\hat{Q}_2 ).
\label{eq:4.5}
\end{equation}

As in the previous section, the Hartree approximation of any generator, the
corresponding classical variable, will be denoted by the same symbol without
a hat, the map of a product will be the product of the maps, but the map of
a commutator will be i times the Poisson bracket of the two maps in the same
order.  We can now state the theorem that generalizes the one for $n=1$. \\
{\em Theorem}:  For a class of many-Fermion Hamiltonians belonging to the
enveloping algebra of Sp(2,R), of which Eq.\ (\ref{eq:4.5}) is a prototype,
there exists a six-dimensional family of Slater determinants, defined by
density matrices, $\rho({\bf x},{\bf x}'|Q_0 ,P_0 ,Q_1 ,P_1 ,Q_2 ,P_2 )$,
that describe states belonging to an irreducible representation, which\\
(i) induce a mapping of Sp(2,R) onto a symplectic manifold $(Q_0,P_0,Q_1,P_1,
Q_2,P_2) \equiv (Q,P)$
in which the generators, $\hat{G}$, are mapped as classical dynamical
variables, i.\ e.\ ,       \begin{equation}
\hat{G} \rightarrow  G(Q,P).    \label{eq:4.6}     \end{equation}
The mapping of products and commutators is as previously specified. \\
(ii) The density matrices have the diagonal form   \begin{equation}
\rho(x,x'|Q,P) = \sum_h \psi_h(x|Q,P)\psi^{\ast}_h(x'|Q,P), \label{eq:4.7}
\end{equation}
where the sum is over occupied orbitals, $h$.  Furthermore $\psi_h (x|QP)$
has the form   \begin{eqnarray}
\psi_h (x|Q,P) &=& \exp[i{\cal S}(x,P)] \phi_h (x,Q),  \label{eq:4.8}  \\
{\cal S}(x,P) &=& \mbox{$\frac{1}{2}$} (x_1^2 +x_2^2)P_0 + \mbox{$\frac{1}{2}$}
(x_1^2 -x_2^2)P_1 \nonumber\\&&
+x_1 x_2 P_2,
\label{eq:4.9}   \end{eqnarray}
and $\phi_h$ is the solution of a constrained Hartree equation that will
be specified and solved in the course of the demonstration.  Altogether,
these results will establish that the density matrix (\ref{eq:4.7}) is
a solution of the time-dependent Hartree equation in the density-matrix
form.

We turn to the proof of this theorem.  Since the technique is a relatively
straightforward generalization of the proof given in the previous section,
we shall be sparing of details, except in so far as these bring
in something novel compared to the previous case.  The first step is to replace
the commutator algebra by a Poisson-bracket (PB) algebra.
One new aspect is that we
have an angular-momentum generator, $L_{12}$. The PB relations between the
coordinates and this quantity determine it to have the (not surprising) value
\begin{equation}  L_{12}=Q_1P_2 -Q_2P_1. \label{eq:4.10}   \end{equation}
Next we consider the PB relations of the coordinates with the
``deformation" generators, $S_{ij}$.  We thereby
obtain nine differential equations
which have the solutions,           \begin{eqnarray}
S_{11} &=& (P_1 +P_0)(Q_1 +Q_0) +P_2 Q_2,      \label{eq:4.11}  \\
S_{22} &=& (P_1 -P_0)(Q_1 -Q_0) +P_2 Q_2,     \label{eq:4.12}    \\
S_{12} &=& Q_0 P_2 +Q_2 P_0.     \label{eq:4.13}     \end{eqnarray}
A simple check on these results is that
the sum of $S_{11}$ and $S_{22}$ must be
a rotational invariant.
We find   \begin{equation}
S_{11}+S_{22} = 2P_0 Q_0 +2P_1 Q_1 +2P_2 Q_2. \label{eq:4.14} \end{equation}

With the above results, we can determine the momentum generators, $K_{ij}$,
by considering the nine PB relations of the coordinates with these
generators, which are determined by the known values for $S_{ij}$ and $L_{12}$.
We thus find       \begin{eqnarray}
P_{11}&=&(Q_0 +Q_1 )(P_0 +P_1 )^2 +(Q_0 -Q_1 )P_2^2
\nonumber\\&&
+2Q_2 P_2 (P_0 +P_1 )
+\chi_{11}(Q),  \label{eq:4.15}   \\
P_{22}&=&(Q_0 -Q_1 )(P_0 -P_1 )^2 +(Q_0 +Q_1 )P_2^2
\nonumber\\&&
+2Q_2 P_2 (P_0 -P_1 )
+\chi_{22}(Q),  \label{eq:4.16}    \\
P_{12}&=&2Q_0 P_0 P_2 +2Q_1 P_1 P_2 +Q_2 (P_0^2 -P_1^2 +P_2^2 ) \nonumber\\&&
+\chi_{12}(Q).
\label{eq:4.17}    \end{eqnarray}
Here, $\chi_{ij}(Q)$ are three unknown functions of $Q$ that remain to be
determined for a completion of the mapping.  Notice in passing that the
classical kinetic energy,
\begin{eqnarray}
K& =& \mbox{$\frac{1}{2}$} (P_{11} +P_{22}) =Q_0 (P_0^2 +P_1^2 +P_2^2 )
\nonumber\\&&
+2P_0 (Q_1 P_1 +Q_2 P_2
)+ \mbox{$\frac{1}{2}$} (\chi_{11} +\chi_{22}),     \label{eq:4.18}
\end{eqnarray}
is rotationally invariant, provided the sum of the last two terms on the
right hand side has this property.

For the determination of the unknown functions, $\chi_{ij}$, we study the
PB relations of the deformation generators $S_{ij}$ with the momentum
generators $K_{ij}$. As an example, the three relations involving $S_{ij}$
and $P_{12}$ yield the differential equations,      \begin{eqnarray}
(Q_0 +Q_1 )(\frac{\partial \chi_{12}}{\partial Q_0} +\frac{\partial \chi_{12}}
{\partial Q_1}) +Q_2\frac{\partial \chi_{12}}{\partial Q_2} &=& -\chi_{12},
\label{eq:4.19} \\
(Q_1 -Q_0 )(\frac{\partial \chi_{12}}{\partial Q_0} -\frac{\partial \chi_{12}}
{\partial Q_1}) -Q_2\frac{\partial \chi_{12}}{\partial Q_2} &=& \chi_{12},
\label{eq:4.20}   \\
-Q_2\frac{\partial \chi_{12}}{\partial Q_0} -Q_0\frac{\partial \chi_{12}}
{\partial Q_2} &=&
\chi,
\label{eq:4.21}
\end{eqnarray}
where $\chi = \mbox{$\frac{1}{2}$} (\chi_{11} +\chi_{22})$.
These equations have the solutions  \begin{eqnarray}
\chi_{12} &=& \frac{CQ_2}{Q_0^2 -Q_1^2 -Q_2^2},  \label{eq:4.22} \\
\chi&=& \frac{-CQ_0}{Q_0^2 -Q_1^2
-Q_2^2},   \label{eq:4.23}    \end{eqnarray}
whereas the remaining differential equations are compatible with the above
and in addition yield        \begin{equation}
\chi_- \equiv \mbox{$\frac{1}{2}$} (\chi_{11} -\chi_{22}) = \frac{CQ_1}{Q_0^2
-Q_1^2 -Q_2^2}.
\label{eq:4.24}  \end{equation}

In order, finally, to determine the constant $C$, we calculate the map,
${\cal C}_2$, of the second order Casimir invariant in two ways.  On the
one hand we substitute the maps of the individual generators into the
appropriate expression and thus find   \begin{equation}
{\cal C}_2 = 2C.    \label{eq:4.25}    \end{equation}
On the other hand, we calculate the Hartree value of the invariant directly,
as done in Appendix A.  We thereby obtain (for a special
representation)   \begin{equation}
C= \frac{2}{9} N^3.     \label{eq:4.26}    \end{equation}

We complete part (i) of this demonstration by displaying the collective
Hamiltonian that emerges from these considerations,   \begin{eqnarray}
H_C &=& Q_0 (P_0^2 +P_1^2 +P_2^2 ) +2P_0 (Q_1 P_1 +Q_2 P_2 )
\nonumber\\&&
+V(Q),
\label{eq:4.27}  \\
V(Q) &=& Q_0 +\mbox{$\frac{1}{2}$}\kappa_0 Q_0^2 +\mbox{$\frac{1}{2}$}\kappa_2
(Q_1^2 +Q_2^2 )
+\frac{CQ_0}{Q_0^2 -Q_1^2 -Q_2^2}. \nonumber\\&&
\label{eq:4.28}  \end{eqnarray}

For part (ii) of the theorem, we must construct the density matrix that
solves the time-dependent Hartree equation.  The first step is to separate
$\psi(x|Q,P)$ into two factors, as done in Eq.\ (\ref{eq:4.8}).
{}From the time derivatives
of the coordinates, calculated in two equivalent ways, from the classical
equations of motion and from the Hartree averages of the quantum equations
of motion, we obtain three equations, one for each coordinate,
that generalize the
single equation obtained by combining (\ref{eq:2.49}) and (\ref{eq:2.50}).
These yield the solution
\begin{equation}
{\cal S}(x,P) = \mbox{$\frac{1}{2}$} (x_1^2 +x_2^2 )P_0 +\mbox{$\frac{1}{2}$}
(x_1^2 -x_2^2 )P_1
+x_1 x_2 P_2.     \label{eq:4.29}     \end{equation}

In analogy with Eqs.\ (\ref{eq:2.52}) and (\ref{eq:2.53}),
we introduce the orbital that is the solution to the time-dependent
Hartree equation, $\tilde{\psi}_h$,
leading to the equation    \begin{equation}
\epsilon_h\psi_h +i\sum_{\mu=0,1,2}(\dot{P}_\mu\frac{\partial \psi_h}
{\partial P_\mu} +\dot{Q}_\mu\frac{\partial \psi_h}{\partial Q_\mu})
={\cal H}\psi_h,  \label{eq:4.30}     \end{equation}
where the time derivatives are to be replaced by the classical equations of
motion and the Hartree Hamiltonian, ${\cal H}$, has the form   \begin{equation}
{\cal H} = k_0 +q_0 +\kappa_0 q_0 Q_0 +\kappa_2 (q_1 Q_1 + q_2 Q_2).
\label{eq:4.31}   \end{equation}
As before terms of second order in the classical momentum variables cancel.
The zero-order terms yield the constrained Hartree equation, \begin{eqnarray}
\epsilon_h\phi_h &=& ({\cal H} - \sum \lambda_\mu q_\mu)\phi_h,
\label{eq:4.32}  \\
\lambda_\mu &=& \frac{\partial V}{\partial Q_\mu}.  \label{eq:4.33}
\end{eqnarray}      The first-order terms yield three equations
that state, in analogy with Eq.\ (\ref{eq:2.56}), the
identity, when acting on $\phi_h$,
between certain linear
operators in the collective coordinates and corresponding linear operators
in the single-particle coordinates.  The
statement and proof of these identities is given
in Appendix C in order not to interfere with the flow of the main argument.

We turn then to the solution of the constrained Hartree equation,
(\ref{eq:4.32}).
The operator that appears on the right hand side of this equation
will be called ${\bar{\cal H}}$.  The cross-terms in the potential energy
can be eliminated by an orthogonal transformation to the intrinsic
system,   \begin{eqnarray}
x_1 &=& \cos\theta \bar{x}_1 +\sin\theta \bar{x}_2,   \label{eq:4.34} \\
x_2 &=& -\sin\theta\bar{x}_1 +\cos\theta\bar{x}_2,   \label{eq:4.35} \\
\cos 2\theta &=& -(\tilde{\omega}_1^2 -\tilde{\omega}_2^2 )/D_1,
\label{eq:4.36}  \\
\sin 2\theta &=& -4\lambda_{12}/D_1, \label{eq:4.37}  \\
\tilde{\omega}_1^2 &=& \frac{C}{D_0} -\frac{CQ_0 (Q_0 -Q_1 )}{D_0^2},
\label{eq:4.38} \\
\tilde{\omega}_2^2 &=& \frac{C}{D_0} -\frac{CQ_0 (Q_0 +Q_1 )}{D_0^2},
\label{eq:4.39} \\
\lambda_{12} &=& -\frac{CQ_0 Q_2}{D_0^2}, \label{eq:4.40} \\
D_0 &=& Q_0^2 -Q_1^2 -Q_2^2 , \label{eq:4.41} \\
D_1 &=& |\sqrt{16\lambda_{12}^2 +(\tilde{\omega}_1^2 -\tilde{\omega}_2^2)^2}|.
\label{eq:4.42a}    \end{eqnarray}
These equations transform $\bar{{\cal H}}$ to the form  \begin{equation}
\bar{{\cal H}} = \mbox{$\frac{1}{2}$} (\bar{p}_1^2 +\bar{\omega}_1^2 x_1^2 )
+\mbox{$\frac{1}{2}$} (\bar{p}_2^2 +\bar{\omega}_2^2 \bar{x}_2^2 ),
\label{eq:4.42}
\end{equation}    where \begin{eqnarray}
\bar{\omega}_1^2 &=& |C|/(Q_0 +Q)^2,  \label{eq:4.43}  \\
\bar{\omega}_2^2 &=& |C|/(Q_0 -Q)^2, \label{eq:4.44} \\
Q &=& \sqrt{Q_1^2 +Q_2^2} = \bar{Q}_1.   \label{eq:4.45}    \end{eqnarray}
The last of these expressions shows that the  intrinsic system is the one
in which $\bar{Q}_2$ vanishes.

In consequence of the above transformation, the solution of the constrained
Hartree equation is   \begin{equation}
\phi_h (x,Q) = (\bar{\omega}_1\bar{\omega}_2 )^{1/4}\phi_{h_1}^{(sho)}(
\sqrt{\bar{\omega}_1}\bar{x}_1 )\phi_{h_2}^{(sho)}(\sqrt{\bar{\omega}_2}
\bar{x}_2 ).   \label{eq:4.46}     \end{equation}
Together with the material relegated to Appendix C this completes the proof
of the theorem.

\subsection{Application of the theory of large amplitude collective
motion}

Following the procedure outlined for the one-dimensional case,  the first
step is to compute the collective potential energy.  Because in the intrinsic
frame the Hartree Hamiltonian is a sum of two harmonic-oscillator
contributions (with different frequencies), the procedure, including the
use of the virial theorem, for obtaining a differential equation for the
potential energy follows through without a hitch.  Here it yields the
partial differential equation    \begin{equation}
V=2Q_0 +\frac{3}{2}(\kappa_0 Q_0^2 +\kappa_2 \bar{Q}_1^2 )
-Q_0\frac{\partial V}{\partial Q_0} -\bar{Q}_1\frac{\partial V}{\partial
\bar{Q}_1},  \label{eq:4.47}   \end{equation}
with the solution   \begin{equation}
V = Q_0 + \mbox{$\frac{1}{2}$} (\kappa_0 Q_0^2 +\kappa_2\bar{Q}_1^2 )
+\frac{|C|Q_0}
{Q_0^2 -\bar{Q}_1^2}.  \label{eq:4.48}   \end{equation}
In a general coordinate system, we should replace
$\bar{Q}_1$ by $\sqrt{Q_1^2 + Q_2^2}$.  Thus we recognize also in the
present context that $V$ is a scalar.

Consider next the kinetic energy, in the form  \begin{equation}
T = \mbox{$\frac{1}{2}$} B^{\mu\nu}P_\mu P_\nu,   \label{eq:4.49}
\end{equation}
where the indices take the values $0,1,2$.  We illustrate the two steps that
enter into the calculation of the mass matrix elements $B^{\mu\nu}$.
As one of three equations for the time derivatives of the coordinates,
calculated in two ways, we have   \begin{equation}
\dot{Q}_1 = S_{11}-S_{22}=\frac{\partial H_C}{\partial P_1}
=B^{1\mu}P_\mu.   \label{eq:4.50}   \end{equation}
As one of six equations necessary to complete the calculation, we have
\begin{equation}
[Q_1 ,S_{11}-S_{22}]=\frac{\partial (S_{11}-S_{22})}{\partial P_1}
=2Q_0 =B^{11},   \label{eq:4.51}    \end{equation}
that combines a known PB relation with a derivative of (\ref{eq:4.50}).
The completion of the procedure just exemplified yields the mass matrix
previously determined by the theorem, that we display in matrix form,
\begin{equation}
{\bf B}=\left(\begin{array}{ccc} 2Q_0 & 2Q_1 & 2Q_2 \\ 2Q_1 & 2Q_0 & 0 \\
2Q_2 & 0 & 2Q_0 \end{array} \right).    \end{equation}
As explained in the previous section, this establishes once more
that we have an exactly decoupled manifold.
Again the alternative calculation within the framework of our theory
of collective motion is summarized in Appendix B.

\section{Monopole plus quadrupole model associated with the algebra of
Sp(3,R)}
\subsection{Classical mapping and Hartree solution}

In this section, that indeed quotes the results of most future interest to
us, we shall drop the pretensions of formality adhered to in the previous
sections.  Except for very few points emphasized below, the results to be
proved as well as the techniques used to carry out the demonstrations
should be evident by now.
We divide the  generators into monopole and quadrupole parts.
In order to keep better track of the significance of the variables, we adopt an
alphanumeric subscript notation, illustrated by means of the one-particle
operators,          \begin{eqnarray}
q_s & = & \frac{1}{3}(x_1^2+x_2^2+x_3^2), \\
q_{d1} & = & -\frac{1}{3}(x_1^2+x_2^2-2x_3^2) ,\\
q_{d2} & = & \frac{1}{2}(x_1^2-x_2^2),\\
q_{o1} & = & x_1x_2,\\
q_{o2} & = & x_1x_3,\\
q_{o3} & = & x_2x_3,\\
s_s & = & \frac{1}{3}(x_1p_1+x_2p_2+x_3p_3) ,\\
s_{d1} & = & -\frac{1}{3}(x_1p_1+x_2p_2-2x_3p_3) ,\\
{\rm etc.},\nonumber\\
k_s & = & \frac{1}{3}(p_1^2+p_2^2+p_3^2) ,\\
k_{d1} & = & -\frac{1}{3}(p_1^2+p_2^2-2p_3^2) ,\\
{\rm etc.} \nonumber
\end{eqnarray}

Associated with these one-particle operators are second quantized operators,
designated by the same symbols in the upper case and carrying hats, and the
Hartree maps of these, designated by the uppercase but hatless.  By means
of the PB algebra these are expressed as functions of six canonical pairs,
$(Q_s,P_s ),\cdots ,(Q_{o3},P_{o3})$.  The resulting espressions, which were
calculated with the aid of the program, Mathematica,  are too long to quote
in their raw form.  To obtain more concise expressions, we revert to a
spherical
tensor notation
\begin{eqnarray}
\hat{Q}_0    & = & \frac{\sqrt{3}}{2} \hat Q_{d1},\\
\hat{Q}_1    & = & -\frac{1}{\sqrt{2}} (\hat Q_{o2} + i \hat{Q}_{o3}),\\
\hat{Q}_{-1} & = & \frac{1}{\sqrt{2}} (\hat Q_{o2} - i \hat{Q}_{o3}),\\
\hat{Q}_2    & = & \frac{1}{\sqrt{2}} (\hat Q_{d2} +i \hat{Q}_{o1}),\\
\hat{Q}_{-2} & = & \frac{1}{\sqrt{2}} (\hat Q_{d2} -i \hat{Q}_{o1}),
\end{eqnarray}
for $Q$, and similar definitions for $S$ and $K$.  In the following
expressions,
the classical limit of $\hat Q_m$ is denoted by $\Xi_m$, and the conjugate
momenta are denoted by $\Pi_m$. We use $Q_s$ and $P_s$ for the monopole
coordinate and its momentum. The spherical components of $S$ are denoted
by $\Sigma$, etc..  Finally, $L_m$ will refer below to the standard
spherical components of the angular momentum vector.

We thus find that
\begin{eqnarray}
L_m & = & i\sqrt{10}\left[\Xi\times\Pi\right]^{(1)}_m, \\
S_s & = & \frac{2}{3} (\Xi\cdot \Pi + Q_s P_s),\\
\Sigma_m & = & -\sqrt{\frac{7}{6}} \left[\Xi\times\Pi\right]^{(2)}_m
+\frac{2}{3}P_s\, \Xi_m + Q_s \Pi_m,\\
K_s & = &
-\frac{\sqrt{42}}{9}\left[\Pi\times\Pi\right]^{(2)}\cdot\Xi
+ \frac{8}{9}P_s\, \Xi \cdot \Pi +\frac{4}{3} Q_s\, \Pi \cdot \Pi
\nonumber\\&&
+ \frac{4}{9} Q_sP_s^2+\chi_s,\\
K_m & = & \frac{\sqrt{5}}{15}
\left[\left[\Pi\times\Pi\right]^{(0)}\times\Xi\right]^{(2)}_m
-\frac{1}{3}
\left[\left[\Pi\times\Pi\right]^{(2)}\times\Xi\right]^{(2)}_m
\nonumber\\&&
+\frac{6\sqrt{5}}{5}
\left[\left[\Pi\times\Pi\right]^{(4)}\times\Xi\right]^{(2)}_m
-\frac{2\sqrt{42}}{9}P_s \left[\Xi\times\Pi\right]^{(2)}_m
\nonumber\\&&
+\frac{4}{9} P_s^2\, \Xi_m +X_m ,\\
\chi_s & = & -C \partial_{Q_s} \ln\left(4Q_s^3-4\Xi\cdot\Xi\, Q_s
-I_3(\Xi)\right),  \label{eq:5.21} \\
X_m & = & -\frac{3}{2} C \partial_{\Xi_{-m}}\ln\left(4Q_s^3-4\Xi\cdot\Xi\, Q_s
-I_3(\Xi)\right),\\
I_3(\Xi) & =& \frac{8}{3\sqrt{3}}\sqrt{\frac{35}{2}}\left[\left[\Xi\times
\Xi\right]^{(2)}\times\Xi\right]^{(0)}, \\
{\cal C}_2 & = &-3 \phi_s Q_S -2\sum_m \Phi_m \Xi_m = 9 C,\\
H_C & = & \frac{3}{2} (K_s+Q_s) +\mbox{$\frac{1}{2}$} \kappa_0 Q_s^2+
\mbox{$\frac{1}{2}$} \kappa_2 \Xi\cdot\Xi +\mbox{$\frac{1}{2}$}
\kappa_L L\cdot L.\nonumber\\&&
\end{eqnarray}
The above expressions contain the standard notation for angular momentum
coupling.
We recognize $\chi_s$
and $X_m$ as the singular functions that arise in the determination of the
momentum generators.  Here the partial signs represent derivatives with
respect to the variables written, and $C$ is the integration constant that has
appeared analogously in the calculations for lower dimensionality. A
complication that has {\em not} appeared previously is the occurrence of the
cubic invariant, $I_3$, constructed from the tensor $\Xi_m$.
Finally, from the given form of the collective Hamiltonian,
$H_C$, it is obvious what
starting many-body Hamiltonian was used.  Compared to the corresponding
two-dimensional case, we have only added an angular momentum coupling.

To construct the density matrix, we write     \begin{equation}
\psi_h(x|Q,P) = \exp[i{\cal S}(x,P)]\phi_h(x,Q),  \end{equation}
and find    \begin{equation}
{\cal S}(x,P) = P_s q_s + \Pi\cdot q.   \end{equation}
For the real orbitals, $\phi_h$, we find   \begin{equation}
\phi_h = \prod_{i=1}^3 (\bar{\omega}_i)^{1/4}\phi_{h_i}^{(sho)}(\sqrt{\bar{
\omega}_i}\bar{x}_i),   \end{equation}
as the solution of a constrained Hartree equation, referred to the intrinsic
or barred system (Cf.\ (\ref{eq:4.43})).  For the constrained Hamiltonian
we write  \begin{eqnarray}
\bar{{\cal H}} &=& \mbox{$\frac{1}{2}$}\sum_i (\bar{p}_i^2 +\bar{x}_i^2 )
+q_s(\kappa_0 \bar{Q}_s
-\frac{\partial V}{\partial\bar{Q}_s})     \nonumber \\
&& +q_0(\kappa_2\bar{Q}_0 -\frac{\partial V}{\partial\bar{Q}_0})
+q_{d_2}(\kappa_2 \bar{Q}_{d_2}-\frac{\partial V}{\partial\bar{Q}_{d_2}}) \\
&=& \mbox{$\frac{1}{2}$} \sum_i (\bar{p}_i^2 +\bar{\omega}_i^2 \bar{x}_i^2 ),
\\
\bar{\omega}_1^2 &=& 1+\frac{2}{3}(\kappa_0 \bar{Q}_s -\frac{\partial V}
{\partial\bar{Q}_s})  \nonumber \\
&&-\frac{1}{\sqrt{3}}(\kappa_2\bar{Q}_0 -\frac{\partial V}{\partial\bar{Q}_0})
+(\kappa_2 \bar{Q}_{d_2} -\frac{\partial V}{\partial\bar{Q}_{d_2}}), \\
\bar{\omega}_2^2 &=& 1+\frac{2}{3}(\kappa_0 \bar{Q}_s -\frac{\partial V}
{\partial\bar{Q}_s})  \nonumber \\
&&-\frac{1}{\sqrt{3}}(\kappa_2\bar{Q}_0 -\frac{\partial V}{\partial\bar{Q}_0})
-(\kappa_2 \bar{Q}_{d_2} -\frac{\partial V}{\partial\bar{Q}_{d_2}}), \\
\bar{\omega}_3^2 &=& 1+\frac{2}{3}(\kappa_0 \bar{Q}_s -\frac{\partial V}
{\partial\bar{Q}_s})  \nonumber \\
&&+\frac{2}{\sqrt{3}}(\kappa_2\bar{Q}_0 -\frac{\partial V}{\partial\bar{Q}_0}).
\end{eqnarray}
Alternative expressions for the frequencies that may be found by substituting
the derivatives of the potential energy will not be recorded here.

In addition to the above relations and conditions, there are also equations
that relate certain linear differential operators with respect to the
collective coordinates to linear differential operators with respect to
the intrinsic coordinates.  These were discussed for the two-dimensional
case in Appendix C, but will not be discussed at all for this case.

\subsection{Application of the theory of large amplitude collective motion}

Following the procedure developed for the one-dimensional case and previously
also applied to the two-dimensional case, we obtain a partial
differential equation for the potential energy.  In stating this equation
and its solution we shall suppress the bar notation, understanding that
the collective coordinates refer to the intrinsic system. We thus find
\begin{eqnarray}
V&=& 3Q_s +\frac{3}{2}[\kappa_0 Q_s^2 +\kappa_2 (Q_0^2 + Q_{d_2}^2 )] \nonumber
\\  && -Q_s\frac{\partial V}{\partial Q_s} -Q_0\frac{\partial V}{\partial Q_0}
-Q_{d_2}\frac{\partial V}{\partial Q_{d_2}}.  \end{eqnarray}
This equation has the solution    \begin{equation}
V= \frac{3}{2}Q_s +\mbox{$\frac{1}{2}$} \kappa_0 Q_s^2 +\mbox{$\frac{1}{2}$}
\kappa_2 (Q_0^2 +Q_{d_2}^2)
+ \chi_s,    \end{equation}
where $\chi_s$ is the singular scalar function displayed in Eq.\
(\ref{eq:5.21}).  In the present calculation, the quadratic and cubic
scalars, $I_2$ and $I_3$ are found in the versions to which they reduce
in the intrinsic system, namely,
\begin{eqnarray}
I_2 &=& Q_0^2 +Q_{d_2}^2,    \\
I_3 &=& -\frac{8}{3\sqrt{3}}(Q_0^3 -3Q_0 Q_{d_2}^2).      \end{eqnarray}

There remains only the problem of computing the mass tensor.  This remains
a simple algebraic task, if one applies the method first described
for the one-dimensional case.  The calculation is simplest to carry out
in terms of Cartesian variables, but the equivalence of the result to that
given by Eqs.\ (5.19) and (5.25) can then be verified.
Nothing new is learned by repeating
either the calculation or the results.

\section{Concluding remarks}

In this paper we have developed a new exactly solvable example for the theory
of large amplitude collective motion by looking at the classical limit of
the algebra of Sp(3,R)  and a Hamiltonian defined within the enveloping
algebra of the algebra.  Possibilities exist both for the application and for
theoretical refinement of the results of this paper.  A natural first
application would involve the requantization of the classical collective
Hamiltonian and comparison of the results of diagonalizing the resultant
Bohr Hamiltonian with the results of an exact diagonalization of the
corresponding many-particle Hamiltonian \cite{P}.  A theoretical
refinement would be to attempt to upgrade the classical mapping to a
full quantum boson mapping \cite{K}.  This is probably not difficult for the
one- and two-dimensional cases, but might prove laborious for the interesting
$n=3$ case.

It would be inappropriate to conclude this paper without mentioning possible
connections with other work.  For instance there has been extensive research on
the unitary representations of the non-compact symplectic algebras considered
in this paper \cite{C,R1,R2}, that can be viewed as exact boson mappings of
these algebras.  The relation of our special mapping to this work
may be worth pursuing.  Along a different line, we may be said to have produced
soliton solutions to a highly simplified class of field theories, using
Hamiltonian methods.  It would be interesting to investigate if our methods
can teach us something about the Hamiltonian approach to other soliton
models \cite{F}.

This work was supported in part by the U.\ S.\ Department of Energy
under Grant No.\ 40264-5-25351.

\appendix
\section{Evaluation of second order Casimir invariant}

Consider the one-dimensional case.  The evaluation is based on the idea that
since the monopole-monopole interaction belongs to the enveloping algebra,
changing its strength does not change the irreducible representation.  In
particular, we may set the strength to zero.  In that case the ground state,
that defines the representation of interest to us, is the Slater determinant
obtained by filling the lowest orbitals.  Since in this state the Hartree
value of $\hat{S}$ vanishes, we have from Eq.\ (2.29)  \begin{eqnarray}
{\cal C}_2(1d)
& =& -(\sum_h \epsilon_h)^2 =-(\sum_0^{N-1} n)^2   \nonumber \\
&\cong& \frac{1}{4}N^4.       \end{eqnarray}
In this evaluation, as in the ones that occur below, it is strictly the
definition of the classical variables as Hartree averages that intervene.
Furthermore, we have utilized the virial theorem for the harmonic
oscillator described by the Hamiltonian $\mbox{$\frac{1}{2}$} (p^2 +x^2 )$.

For the two and three-dimensional cases, we shall evaluate the
Casimir invariant only for the closed-shell and for the representation
containing the ground state.  This allows us to take fullest advantage of
circular and spherical symmetry, respectively.  We provide a few details for
three-dimensions only.  We have   \begin{eqnarray}
{\cal C}_2(3d) &=&-\sum_{i=1}^3 Q_{ii}P_{ii}   \nonumber \\
&=& -\frac{1}{3}(\sum_i Q_{ii})^2.     \end{eqnarray}
For $N$ particles, this becomes   \begin{eqnarray}
{\cal C}_2(3d) &=&-\frac{1}{3}(\sum_0^{(6N)^{1/3}}\mbox{$\frac{1}{2}$} n^3)^2
\nonumber \\
&=&-\frac{3}{16}6^{2/3}N^{8/3}.    \end{eqnarray}
Here we have used the fact that the level with energy $n$
has degeneracy $\mbox{$\frac{1}{2}$} n^2$
and that the number of shells needed for $N$ particles is $(6N)^{1/3}$.  These,
of course, are approximate values needed to get the answer to leading order
in $N$.

The corresponding calculation for two dimensions yields \begin{equation}
{\cal C}_2(2d) = -\frac{4}{9} N^3.     \end{equation}
In future applications, we shall be interested in irreducible representations
associated with open-shell nuclei.  The generalization
to such cases of the elementary
calculations just presented will be deferred to the
occassion when those applications  are presented.

\section{Alternative computation of the mass tensor}

In the body of the text we have described a new method for computing
the mass tensor of a decoupled or approximately decoupled manifold.
This method is particularly convenient when the cranking operator or
operators can be represented in a basis of single-particle operators
that depend only on the coordinates, though it can also be applied with
somewhat increased effort in more general cases.  Nevertheless, it is important
to show how the same results can be obtained by the methods developed in
our previous work.  In fact, the monopole model was instrumental in leading
to a necessary generalization of the theory of the mass tensor  as it had
been presented and utilized by us.
The generalization required is presented in some detail in
\cite{ACM1}.  We do not wish to repeat the technical details here, but rather
we shall ``remind" the reader
of the basic idea and be content with the quotation and application of
the final result.

The theory of large amplitude collective motion is based on the idea that
starting from a Hamiltonian, quadratic in the momenta but otherwise
arbitrary, one can introduce a point transformation chosen so that in the
new coordinates the existence of a decoupled coordinate manifold is either
apparent, or in more realistic cases can be demonstrated to be approximately
true.  The restriction to a coordinate or point transformation was based
on the observation that this type of transformation maintained exactly
the quadratic dependence on the momenta.  However, the formulas derived
from this procedure failed to reproduce the correct mass in the monopole
example.  Thereafter, it was discovered \cite{ACM1} that for consistency
to second order in the momentum one had to replace point transformations
by more general canonical transformations, correct to second order in the
momenta.  In the paper cited details were supplied only for the case of
one collective coordinate, but for the separable Hamiltonians considered
in the present paper it is straightforward to generalize the result to
any number of collective coordinates.  Below we describe the result of this
generalization.

We suppose that there is a set of self-consistent cranking operators,
\begin{equation}
Q^i = {\rm tr}(q^i\rho).       \end{equation}
It follows then that the component $B^{ij}$ of the mass tensor is given by
the formula      \begin{eqnarray}
B^{ij} &=& [q^{(i)}_{hp}\bar{{\cal H}}_{pp'}q^{(j)}_{p'h}
            + i\leftrightarrow j]  \nonumber    \\
     && -[q^{(i)}_{ph}\bar{{\cal H}}_{hh'}q^{j}_{h'p}
	    + i\leftrightarrow j],   \\
\bar{{\cal H}}_{ab} &=& {\cal H}_{ab} -\frac{\partial V}{\partial Q^i}
q^{(i)}_{ab}.       \end{eqnarray}
Remarkably, where the present formula contains the cranked Hartree
Hamiltonian, $\bar{{\cal H}}$, the previous, incorrect formula contains
just the unconstrained Hartree Hamiltonian ${\cal H}$.  Precisely because
the single-particle orbits referred to in the above formula are the
eigenmodes of the constrained operator, and using completeness as well,
the formula for the mass tensor can be transformed into the simple
formula involving double commutators,  \begin{equation}
B^{ij} = \mbox{$\frac{1}{2}$} {\rm tr}\{\rho[[q^i ,\bar{{\cal H}}],q^j ]
+i\leftrightarrow j
\}.      \end{equation}

For the cases treated in the text, the $q^i$ are coordinate operators.
Consequently $\bar{{\cal H}}$ may be replaced by the single-particle
kinetic energy.  It is then a trivial calculation to show that all the
results of the text are reproduced.

\section{Some additional details concerning the solution of the
time-dependent Hartree equation}

In the construction of the density matrix that solves the time-dependent
Hartree equation for the two-dimensional case, we bypassed the study of a
set of conditions that must be satisfied by the real orbitals $\phi_h$
in addition to the constrained Hartree equation (\ref{eq:4.32}).  These
conditions, here three in number, arise from terms linear in the collective
momenta in the time-dependent Hartree equation.  They are of the form,
\widetext
\begin{eqnarray}
2Q_0\frac{\partial \phi}{\partial Q_0} +2Q_1\frac{\partial \phi}{\partial Q_1}
+2Q_2\frac{\partial \phi}{\partial Q_2}&=& -x_1\frac{\partial \phi}
{\partial x_1} -x_2\frac{\partial \phi}{\partial x_2} -\phi , \label{eq:C.1}\\
2Q_1\frac{\partial \phi}{\partial Q_0} +2Q_0\frac{\partial \phi}{\partial Q_1}
&=& -x_1\frac{\partial \phi}
{\partial x_1} +x_2\frac{\partial \phi}{\partial x_2} , \label{eq:C.2}\\
2Q_2\frac{\partial \phi}{\partial Q_0} +2Q_0\frac{\partial \phi}{\partial Q_2}
&=& -x_2\frac{\partial \phi}
{\partial x_1} -x_1\frac{\partial \phi}{\partial x_2}. \label{eq:C.3}
\end{eqnarray}
\narrowtext

To show that each orbital, $\phi_h$, given by Eq.\ (\ref{eq:4.32})
satisfies these conditions, it is convenient to transform them to the
intrinsic system.  In addition to the transformation equations (\ref{eq:4.34}),
$\cdots$, we introduce polar coordinates \begin{eqnarray}
Q_1 &=& Q\cos \Theta, \\
Q_2 &=& Q\sin \Theta,   \end{eqnarray}
that are recognized as intrinsic coordinates, if we make the identifications
$Q=\bar{Q}_1 $ and $\Theta = -2\theta$.   Straightforward calculation then
permits us to replace Eqs.\ (\ref{eq:C.1})-- (\ref{eq:C.3}) by the equations
\begin{eqnarray}
2Q_0\frac{\partial \phi}{\partial Q_0} +2Q\frac{\partial \phi}{\partial Q}
&=& -\bar{x}_1\frac{\partial \phi}{\partial \bar{x}_1}
-\bar{x}_2\frac{\partial \phi}{\partial \bar{x}_2} -\phi , \label{eq:C.4}\\
2Q\frac{\partial \phi}{\partial Q_0} +2Q_0\frac{\partial \phi}{\partial Q}
&=& -\bar{x}_1\frac{\partial \phi}{\partial \bar{x}_1}
+\bar{x}_2\frac{\partial \phi}{\partial \bar{x}_2} , \label{eq:C.5}\\
\frac{2Q_0}{Q}\frac{\partial \phi}{\partial \Theta}
&=& -\bar{x}_2\frac{\partial \phi}{\partial \bar{x}_1}
-\bar{x}_1\frac{\partial \phi}{\partial \bar{x}_2}. \label{eq:C.6}
\end{eqnarray}

To verify the last set of equations, we utilize the explicit form
(\ref{eq:4.47}) of the orbitals, as well as the Eqs.\ (\ref{eq:4.44}):
and (\ref{eq:4.45}) that relate the barred frequencies to the collective
coordinates.  We must also remember that the barred coordinates are
also functions of the collective coordinates, according to the equations
that are the inverses of (\ref{eq:4.34}) and (\ref{eq:4.35}) and the
relation between $\theta$ and $\Theta$.  After some algebra, we find that
we can duplicate (\ref{eq:C.4}) and (\ref{eq:C.5}), but in place of
(\ref{eq:C.6}), we deduce the result      \begin{equation}
2\frac{\partial \phi}{\partial \Theta} = -\bar{x}_1\frac{\partial \phi}
{\partial \bar{x}_2} +\bar{x}_2\frac{\partial \phi}{\partial \bar{x}_1},
\label{eq:C.7}      \end{equation}
which obviously describes correctly the rotational properties of an orbital,
in view of the connection between $\Theta$ and $\theta$.
To verify (\ref{eq:C.6}), finally, we calculate the ``commutator" of Eqs.\
(\ref{eq:C.5}) and (\ref{eq:C.7}) and indeed deduce the desired result.

The corresponding calculations for three dimensions will not be reproduced
here.

\end{document}